\documentclass[notoc,preprint,nohyper]{JHEP3}
\usepackage{amssymb}   
\usepackage{amsmath}
\usepackage{mathtools}
\usepackage{makeidx}
\usepackage{graphicx}
\usepackage{axodraw4j}
\usepackage{caption}
\usepackage{subcaption}
\usepackage[utf8]{inputenc}
\usepackage{feynmp-auto}

\usepackage[sort]{cite}

\def\beq{\begin{equation}}
\def\eeq{\end{equation}}
\def\bea{\begin{eqnarray}}
\def\eea{\end{eqnarray}}

\def\f21{{}_2F_{1}}

\def\O{\mathcal{O}}

\DeclareMathOperator{\Li}{Li}

\def\bsp#1\esp{\begin{split}#1\end{split}}
\newcommand{\zb}{\bar{z}}

\catcode`\@=11
\font\manfnt=manfnt
\def\Watchout{\@ifnextchar [{\W@tchout}{\W@tchout[1]}}
\def\W@tchout[#1]{{\manfnt\@tempcnta#1\relax%
  \@whilenum\@tempcnta>\z@\do{%
    \char"7F\hskip 0.3em\advance\@tempcnta\m@ne}}}
\let\foo\W@tchout
\def\dubious{\@ifnextchar[{\@dubious}{\@dubious[1]}}

\def\@dubious[#1]{%
  \setbox\@tempboxa\hbox{\@W@tchout#1}
  \@tempdima\wd\@tempboxa
  \list{}{\leftmargin\@tempdima}\item[\hbox to 0pt{\hss\@W@tchout#1}]}
\def\@W@tchout#1{\W@tchout[#1]}
\catcode`\@=12
\preprint{CERN-TH-2017-090, SLAC-PUB-16962}

\title{Higgs-differential cross section at NNLO in dimensional regularisation}

\author{
Falko Dulat$^a$, Simone Lionetti$^b$, Bernhard Mistlberger$^c$, Andrea Pelloni$^b$, Caterina Specchia$^b$\\
{}$^a$SLAC National Accelerator Laboratory, Stanford University, Stanford, CA 94309, USA\\
{}$^b$Institute for Theoretical Physics, ETH Z\"urich, 8093 Z\"urich, Switzerland\\
{}$^c$CERN Theory Division, CH-1211, Geneva 23, Switzerland\\
}

\abstract{
We present an analytic computation of the Higgs production cross
section in the gluon fusion channel,
which is differential in the components of the Higgs momentum and inclusive in the associated
partonic radiation through NNLO in perturbative QCD. Our computation includes the necessary higher order
terms in the dimensional regulator beyond the finite part that are required for renormalisation
and collinear factorisation at N$^3$LO. We outline in detail the computational methods which we employ.
We present numerical predictions for realistic final state observables, specifically
distributions for the decay products of the Higgs boson in the $\gamma\gamma$ decay channel.
}

\begin{document}

\section{Introduction}

The discovery of the Higgs boson in 2012 at the Large Hadron Collider (LHC) by the ATLAS~\cite{Aad2012} and CMS~\cite{Chatrchyan2012} experiments founded a new era of precision Higgs physics.
The impressive statistical accuracy of the experimental measurements 
has lead to tight constraints on the couplings of Higgs boson interactions. 
The Run 2 of the LHC promises a plethora of further measurements which will allow
to probe the nature of electroweak symmetry breaking with an unprecedented precision.

In order to truly exploit the potential of the LHC, the excellent experimental results must be 
compared to equally precise theoretical predictions. 
The demand for high precision theory has been met in recent years with a flurry 
of calculations at next-to-leading (NLO) and next-to-next-to-leading-order (NNLO) in perturbative QCD.
Higher order corrections are especially important for Higgs phenomenology, 
in part due to the slow convergence of the perturbative expansion in the strong coupling constant 
for the dominant mode of Higgs hadroproduction via gluon fusion. 

The gluon-fusion total cross section has been computed recently through next-to-next-to-next-to-leading order (N$^3$LO)
in perturbative QCD~\cite{Anastasiou2016,Anastasiou2015} in the limit of an infinite top-quark mass. 
At this level of precision, effects that go beyond the leading approximation of the heavy-top quark effective theory
or a treatment in pure QCD, such as quark mass effects or contributions from weak boson loops, are also important.
The most accurate available predictions for these effects have been combined into a single theoretical prediction
for the inclusive Higgs cross section in ref.~\cite{Anastasiou2016}. 

The Higgs cross section is measured in signal-rich regions of phase space which are 
shaped by carefully designed experimental event selections.
Further restrictions on the phase space of the Higgs boson, its decay products and the associated radiation are required by the detector geometry and response.  
Within their defined acceptance, the experiments have superb capabilities to 
measure a multitude of kinematic distributions for the Higgs boson and its decay products in the years to come.
In addition to the total cross section, it is therefore imperative to have precise theoretical predictions for differential cross sections.

Recently, the $pp \to H+1{\,\,\rm jet}$ fully differential cross section
has been computed at NNLO~\cite{Chen2016,Boughezal2015a,Boughezal2015b,Caola2015,Chen2015}.
Combined with the N$^3$LO inclusive cross section, this has allowed to compute the N$^3$LO Higgs cross section 
with a jet-veto~\cite{Banfi2016}.  This is a first example of a differential cross section in gluon-fusion at this perturbative order.
If a fully differential parton-level Monte-Carlo at N$^3$LO  is achieved in the future,
it will become possible to assess the efficiency of the majority of event selection criteria at the same level 
of accuracy in perturbative QCD as the jet-veto efficiency.  
To achieve this goal, one could attempt to generalise any of the available methods at NNLO (sector-decomposition~\cite{Anastasiou:2005cb,Binoth:2000ps,Hepp:1966eg,Roth:1996pd},
slicing~\cite{Boughezal2011,Boughezal2015,Catani2007,Gaunt2015}, subtraction~\cite{Baernreuther2012,Caola2017,i2007,Ridder2005},
reweighting~\cite{Cacciari:2015jma} and other methods~\cite{Anastasiou2010}) to the next perturbative order.
A generalisation of any of these methods would be a formidable task.
An intermediate goal could therefore be to compute first some specific differential distributions of particular importance,
which are the main ingredients for a fully differential N$^3$LO parton-level Monte-Carlo within a slicing method.

The aim of this article is to compute the Higgs cross section fully differential in all components of the Higgs momentum and its decay products,
treating the associated QCD radiation inclusively (integrating over the unrestricted phase space of all partons in the final state) at NNLO. 
In our computation, we maintain the full dependence of the ``Higgs-differential cross section'' on the dimensional regulator.
This is necessary in order to enable the construction of the counter terms for ultraviolet and initial-state collinear divergences at N$^3$LO. 

Our method employs reverse unitarity~\cite{Anastasiou2002,Anastasiou2003,Anastasiou2003a,Anastasiou2003b,Anastasiou2004a}, 
to map the phase space integrations over partonic radiation onto their loop integral duals.
We then reduce the integrals appearing in the partonic cross sections to master integrals using integration-by-part identities~\cite{Tkachov1981,Chetyrkin1981} and the Laporta algorithm~\cite{Laporta:2001dd}.
The master integrals corresponding to the partonic cross sections with two partons in the final state are novel and we evaluate them 
using two methods: the method of differential equations~\cite{Kotikov1991,Gehrmann2000,Henn2013} and a 
direct integration of the angular phase space variables~\cite{Somogyi:2011ir}.
We are able to express all master integrals in terms of hypergeometric functions which are valid at all orders in the dimensional regulator $\epsilon=(4-d)/2$.
These hypergeometric functions can be expanded to practically any order around the limit $\epsilon=0$ in terms of polylogarithms.
As a result we obtain an analytic formula for all required partonic cross sections through NNLO, including the higher order terms in the $\epsilon$ expansion which are needed as input for a future N$^3$LO calculation.

We implement our results in a computer code and use it to obtain predictions at NNLO for various differential distributions 
which are of interest for LHC experiments. In addition to computing the transverse momentum and rapidity distribution for 
the Higgs boson through NNLO,  we analyse various differential properties of the production and subsequent decay of the Higgs to two photons. 

This article is structured as follows. 
In section~\ref{sec:ddxs} we introduce in detail our definition of ``Higgs-differential'' cross sections. 
In section~\ref{sec:PS} we outline how we separate the integration of Higgs boson and QCD radiation degrees of freedom based on phase space factorisation and the reverse unitarity methodology.
In section~\ref{sec:computation} we perform explicitly the computation of the partonic cross sections for Higgs boson production through NNLO in gluon fusion.
Next, we study several key Higgs boson LHC observables in section~\ref{sec:numerics} that have been obtained using a numeric code built upon our analytic results.
Finally, we give our conclusions in section~\ref{sec:conclusions}.

Note that while we only study differential Higgs boson observables in this article, our ``Higgs-differential'' method is not specific to Higgs boson processes.
In fact, it relies solely on the fact that the Standard-Model Higgs boson is a singlet of QCD.
As such, it can also be applied to compute differential distributions for other colourless final states, such as Drell-Yan or diboson production.

\section{Setup for Differential Cross Sections}
\label{sec:ddxs}

We consider the production of a Higgs boson in proton-proton collisions:
\begin{equation}
{\rm Proton}(P_1) + {\rm Proton}(P_2) \to H(p_h) + X,
\end{equation}
where in parentheses we denote the momenta carried by the external particles in the process.
The four-momenta of the protons in the hadronic center of mass frame are given by
\begin{equation}
P_1 = \frac{\sqrt S}{2} \left(1, 0, 0,  1 \right), \quad P_2 = \frac{\sqrt S}{2} \left(1, 0, 0, -1 \right),
\end{equation}
where $\sqrt{S}$ is the collider center of mass energy.
The inclusive hadronic cross section is related to the one for the partonic processes
\begin{equation}
i(p_1) + j(p_2) \to H(p_h) + X,
\end{equation}
with momenta
\begin{equation}
p_1 = x_1 P_1, \quad p_2=x_2 P_2,
\end{equation}
via the factorisation formula
\bea
\sigma_{PP\rightarrow H+X} &=&\sum\limits_{i,j}\int_0^1 dx_1 dx_2 f_i(x_1)f_j(x_2) \hat \sigma_{ij}(S, x_1, x_2,m_h^2),
\eea
In the above, $p_h^2 = m_h^2$ is the invariant mass of the Higgs boson.
The sum over $i$ and $j$ runs over all possible initial state partons.
$f_i(x)$ are the parton distribution functions
and $\hat \sigma_{ij}(S, x_1, x_2,m_h^2)$ are the partonic cross sections.

In this work we are interested in deriving cross sections for observables $\O$
that are sensitive to the momentum of the Higgs boson $p_h$
and do not depend on the details of the additional radiation $X$.
An observable of this type translates into a measurement function $\mathcal{J}_\O(p_h)$
which multiplies the differential cross section giving a weight to every phase space point.
A typical example for $\O$ is a set of kinematic cuts on the Higgs;
in this case $\mathcal{J}_\O(p_h)$ is equal to one if $p_h$ passes these cuts and zero otherwise,
and it can be written as a product of Heaviside $\theta$-functions.
More complicated, experimentally relevant observables
may also be considered as long as they only depend on $p_h$.
For example, $\mathcal{J}_\O$ may be used to weight the production cross section by appropriate Higgs boson decay matrix elements and phase spaces.
We will refer to cross sections for such observables $\O$ that depend on the kinematic details of the production process only through the Higgs four-momentum as Higgs-differential cross sections.

One possible way to parameterise the Higgs momentum is
\begin{equation}
p_h \equiv \left( E,\, p_x,\, p_y,\, p_z\right) =
\left(\sqrt{p_T^2 + m_h^2} \cosh Y,\; p_T \cos\phi,\; p_T \sin\phi,\; \sqrt{p_T^2 + m_h^2} \sinh Y\right),
\end{equation}
where
\[
Y=\frac{1}{2} \log \left( \frac{E+p_z}{E-p_z} \right),
\qquad
p_T=\sqrt{E^2-p_z^2-m_h^2}.
\]
Here $Y$ is the rapidity of the Higgs boson,
$p_T$ is its momentum in the plane which is transverse to the beam axis,
and $\phi$ is the azimuthal angle.
Due to the symmetry of scattering experiments in the azimuthal plane,
partonic cross sections are always independent of $\phi$.
Without loss of generality, we may therefore write Higgs-differential cross sections as
\begin{multline}
\label{eq:xsdiffhad}
\sigma_{PP\rightarrow H+X}\left[ \O \right]
=\sum\limits_{i,j}
\int_{-\infty}^{+\infty}dY \int_0^\infty dp_T^2
\int_0^{2\pi}\frac{d\phi}{2\pi} \int_0^1 dx_1 dx_2 f_i(x_1)f_j(x_2)\\
\times\frac{d^2 \hat \sigma_{ij}}{d Y dp_T^2}(S, x_1,
x_2,m_h^2,Y,p_T^2) \mathcal{J}_\O(Y,p_T^2,\phi,m_h^2),
\end{multline}
where %
$d^2 \hat \sigma_{ij}/d Y dp_T^2$ is the partonic double-differential $p_T$ and rapidity distribution.

Results for the partonic cross sections are naturally expressed
in terms of the kinematics of the partonic processes.
We thus define the partonic center of mass energy
\[
s=x_1 x_2 S=(p_1+p_2)^2
\]
and the ratios
\[
\tau = \frac{m_h^2}{S},
\qquad
z = \frac{m_h^2}{s} = \frac{\tau}{x_1x_2}.
\]
Furthermore, we introduce the two Lorentz-invariant quantities $x$ and $\lambda$
\beq
\lambda \equiv \frac{s-2p_1\cdot p_h}{s-m_h^2},
\qquad
x \equiv \frac{s(p_1+p_2-p_h)^2}{(s-2p_1 \cdot p_h)(s-2p_2 \cdot p_h)},
\eeq
and the shorthand notation
$\bar{z} \equiv 1-z$,
$\bar{\lambda} \equiv 1-\lambda$,
$\bar{x} \equiv 1-x$.
We can express $p_T$ and $Y$ in terms of these two new variables
and the Bjorken-fractions:
\begin{equation}
\label{eq:vardefpty}
Y=\frac{1}{2} \log \left[
\frac{x_1}{x_2}
\frac{1-\frac{\bar z \bar \lambda}{1-\bar z \lambda x}}{1-\bar z \lambda}
\right],
\qquad
p_T^2 = s \frac{\bar z^2  \lambda  \bar \lambda \bar x}{1-\bar z x \lambda}.
\end{equation}
The Higgs-differential cross sections, when cast in terms of these variables, take the form:
\begin{multline}
\label{eq:xsdiffhad2}
\sigma_{PP\rightarrow H+X}\left[\O\right]=
\tau \sum_{i,j} \int_\tau^1 \frac{dz}{z}\int_{\frac{\tau}{z}}^1 \frac{dx_1}{x_1} \int_0^1 dx \int_0^1 d\lambda \int_0^{2\pi}\frac{d\phi}{2\pi}\\
\times f_i(x_1)f_j\left(\frac{\tau}{x_1 z}\right)
\frac{1}{z}\frac{d^2 \hat \sigma_{ij}}{dx\,d\lambda}(z,x,\lambda,m_h^2)
\mathcal{J}_\O(x_1,z,x,\lambda,\phi,m_h^2).
\end{multline}
Note that we absorbed the Jacobian of the variable transformation from $(p_T,Y)$ to $(x,\lambda)$
into the definition of the partonic Higgs-differential cross section.
Having obtained the above formula, we are now ready to compute the cross
section for any hadron collider observable which is differential in
the Higgs boson momentum from a partonic Higgs-differential distribution in $x$, $\lambda$.

\section{The Higgs-Differential Phase Space}
\label{sec:PS}

In the previous section
we parametrised the degrees of freedom corresponding to the momentum of the Higgs boson with variables $z$, $x$, $\lambda$ and $\phi$
defined \emph{ad hoc}.
In the following it will become clear why this particular parametrization of the Higgs boson degrees of freedom is beneficial to our calculation.
This section is divided in three parts.
In the first part we discuss how the phase space of the final state can be separated into a part that depends solely on the Higgs boson kinematics,
and one that captures all dependence on the additional radiation.
We will determine a set of phase space integrals in such a way that we can carry out the integration over the radiation analytically
while remaining differential in the Higgs boson degrees of freedom.
In the second part we discuss the computation of these integrals using reverse unitarity.
Finally, we will explicitly construct the parametrisation of $p_h$ in terms of $z$, $x$, $\lambda$,
and discuss the features of the remaining phase space.

\subsection{Separation of the Phase Space}

The partonic cross section at a given order in QCD perturbation theory is comprised of phase space integrals
over a sum of real and virtual contributions with different multiplicities of partons in the final state.
Schematically, we may write,
\beq
\label{eq:schematicxs}
\hat \sigma_{ij} \sim \sum_m \int d\Phi_{H+m} \mathcal{M}_{ij\to H+X},
\eeq
where $X$ stands for a set of $m$ final state partons.
We are going to examine the integration of a matrix element for a fixed multiplicity of partons over the phase space measure $d\Phi_{H+m}$.
Due to soft and collinear singularities, such integrations are divergent in four dimensions and we regulate them in dimensional regularisation.
We will elaborate on the detailed structure of the partonic matrix elements through NNLO in section~\ref{sec:computation}.

The integration measure for the phase space describing the production of a Higgs boson and $m$ additional partons
with outgoing momenta $k_1,\ldots,k_m$ is given by
\beq
\label{eq:measure}
d\Phi_{H+m} =
\frac{d^dp_h}{(2\pi)^d} (2\pi)\delta_+(p_h^2-m_h^2)
\left[\prod_{i=1}^m \frac{d^dk_i}{(2\pi)^d} (2\pi)\delta_+(k_i^2)\right]
(2\pi)^d \delta^d\bigg(p_1+p_2-p_h-\sum_{i=1}^{m} k_i\bigg),
\eeq
where
\beq
\delta_+(p^2-M^2)=\theta(p^0-M)\delta(p^2-M^2).
\eeq

In order to compute Higgs-differential cross sections
it is natural to separate the integration over the momentum of the Higgs
from the integral over momenta of the final state state partons.
This can be achieved by inserting unity into eq.~\eqref{eq:measure} using
\beq
\label{eq:PSidentity}
1=\int \frac{d^dk}{(2\pi)^d} (2\pi)^d\delta^d\bigg(k-\sum_{i=1}^m k_i\bigg)\int_0^\infty \frac{d\mu^2}{2\pi}(2\pi)\delta_+(k^2-\mu^2).
\eeq
This identity allows us to factor the $H$ plus $m$ parton phase space into the phase space of two massive particles
and the phase space of $m$ massless partons:
\beq
\label{eq:PSFAC}
d\Phi_{H+m}=\int_0^{\infty} \frac{d\mu^2}{2\pi}d\Phi_{HX}(\mu^2) d\Phi_m(\mu^2),
\eeq
where
\begin{align} 
d\Phi_{HX} &=
\label{eq:2mPS}
\frac{d^dp_h}{(2\pi)^d}(2\pi)\delta_+(p_h^2-m_h^2)
\frac{d^dk}{(2\pi)^d}(2\pi)\delta_+(k^2-\mu^2)
(2\pi)^d\delta^d(p_1+p_2-p_h-k), \\
d\Phi_{m} &=
\label{eq:mPS}
\left[\prod_{i=1}^m \frac{d^dk_i}{(2\pi)^d} (2\pi)\delta_+(k_i^2)\right]
(2\pi)^d\delta^d\bigg(k-\sum_{i=1}^m k_i\bigg).
\end{align}
Note that these formulae are valid for an arbitrary number of final state partons $m\ge 0$. 

A key ingredient of our method is that, because of the class of observables we restricted ourselves to,
the integrals over the extra QCD radiation are not constrained by the definition of the measurement.
This allows us to address the integration over $d\Phi_{HX}(\mu^2)$ and $d\Phi_m(\mu^2)$ separately.

\subsection{Final State Radiation and Reverse Unitarity}
\label{sec:RU}

In the present section we address how the analytic integration over additional final state partons can be performed.
In particular, we advocate that the framework of reverse unitarity~\cite{Anastasiou2002,Anastasiou2003,Anastasiou2003a,Anastasiou2003b,Anastasiou2004a}, that has already been successfully used in the computation of the inclusive cross section~\cite{Anastasiou2013a, Anastasiou:2013mca, Anastasiou:2014vaa,Anastasiou:2014lda,Dulat:2014mda,Anastasiou:2015yha,Duhr:2013msa,Duhr:2014nda},
provides a particularly efficient solution to this task.
This framework exploits the duality between inclusive phase space integrals and loop integrals to treat them in a uniform way. 
Specifically, using Cutkosky's rule~\cite{Cutkosky1960},
it is possible to express the on-shell constraints appearing in phase space integrals through cut propagators
\beq
\delta_+\left(q^2\right)\to\left[\frac{1}{q^2}\right]_\mathrm{c}
= \frac{1}{2\pi i}\text{Disc}\left(\frac{1}{q^2}\right)
= \frac{1}{2\pi i}\left(\frac{1}{q^2+i0}-\frac{1}{q^2-i0}\right).
\eeq
Cut propagators can be differentiated in a similar way to ordinary
propagators with respect to their momenta, 
\begin{equation}
\frac \partial {\partial q_\mu}
\left( \left[ \frac{1}{q^2}\right]_\mathrm{c} \right)^\nu  
= - \nu \left( \left[ \frac{1}{q^2}\right]_\mathrm{c} \right)^{\nu+1} 2 q^\mu\,,  
\end{equation}
leading to identical integration-by-parts (IBP) identities~\cite{Tkachov1981,Chetyrkin1981} for inclusive phase space
integrals as for their dual loop integrals. 
The fact that the cut propagator represents a delta function is reflected by the simplifying constraint
that any integral containing a cut propagator raised to a negative power vanishes:
\beq
\left(\left[\frac{1}{q^2}\right]_\mathrm{c}\right)^{-n} = 0\qquad\text{for}\qquad n \geq 0.
\eeq
IBP identities serve to relate different inclusive phase space and loop integrals
and express them in terms of a finite set of so-called master integrals.
Once a partonic cross section is represented in terms of a linear combination of master integrals,
only those integrals have to be computed by other means. 

In refs.~\cite{Anastasiou2003a,Anastasiou2003b,Anastasiou2004a} a modification of the reverse unitarity framework was developed
in order to allow for the computation of the partonic Drell-Yan and Higgs boson production cross section
differential in the rapidity of the electro-weak final state boson.
Here we want to take this procedure one step further in order to maintain all differential information about the Higgs boson. 
This can be simply achieved by refraining from including the integration over the Higgs boson momentum into the reverse unitarity framework.
Exploiting again the schematic representation of our partonic cross sections eq.~\eqref{eq:schematicxs} we may write
\beq
\label{eq:schematicRU}
\hat \sigma_{ij}(S, x_1, x_2, m_h^2) \sim \sum_m \int_0^{\infty} \frac{d\mu^2}{2\pi}d\Phi_{HX} \times
\underbrace{\left[ \int d\Phi_m \mathcal{M}_{ij\to H+X}\right]}_{\text{Reverse Unitarity}}.
\eeq
To put this into other words: the factorisation of the Higgs boson phase space
and the QCD radiation phase space achieved by eq.~\eqref{eq:PSFAC}
allows us to carry out the integration over the parton momenta separately from the integration over the Higgs boson momentum.
This enables us to use different methods for the two phase spaces without sacrificing any information
about the differential properties of the Higgs boson.
In particular, we can compute the QCD radiation phase space inclusively in dimensional regularisation.
This lends itself naturally to the method of reverse unitarity.
It allows us to apply IBP identities to the 
partonic matrix elements depicted in the square brackets in eq.~\eqref{eq:schematicRU} and express them in terms of differential master integrals. 
These master integrals can subsequently be computed by different means such as direct integration or differential equations, as we will demonstrate below.

The resulting integrated matrix elements are given by a Laurent series in the dimensional regulator $\epsilon$. 
The explicit poles correspond to the regulated soft and collinear singularities of the final state radiation.
In contrast to conventional methods for the computation of differential cross sections,
we circumvent the problem of subtracting infrared singularities among final state partons
by performing these phase space integrals analytically in dimensional regularisation.
We now turn to discuss the specific parametrisation of the remaining degrees of freedom.

\subsection{Parametrising the Higgs Boson Degrees of Freedom}
\label{sec:HiggsPS}
Once the analytic integration over the partonic final state momenta has been performed,
the remaining degrees of freedom are parametrised by the Higgs boson four momentum $p_h$,
or equivalently by the collective momentum of the integrated final state partons $k$.
The inclusive integration over $p_h$ and $k$ is plagued by infrared and collinear singularities
when $k^2=0$ or transverse momenta vanish, $p_{h\;T}^2 = k_T^2 = 0$.
These kinematic limits correspond to the configuration of a single resolved real emission
and to Born level kinematics, respectively.
We will now parametrise the measure of eq.~\eqref{eq:2mPS}
in such a way that leftover divergences can be regulated in a straightforward manner.
We shall see how the parameters $x$ and $\lambda$ introduced in section \ref{sec:ddxs}
are actually a simple solution of this problem.

We parametrise the scalar products of the initial state parton momenta with $k$ by
\begin{equation}
2k\cdot p_1 = \bar{z}\lambda s,
\qquad
2k\cdot p_2 = \bar{z}\tilde{\lambda} s.
\end{equation}
We then find it convenient to exchange the variable $\tilde\lambda$ for $x$ where
\begin{equation}
k^2 = \mu^2 = s \bar{z}^2 \lambda\tilde{\lambda} x,
\qquad
\tilde{\lambda} = \frac{1-\lambda}{1-\bar{z}\lambda x}.
\end{equation}
The parameter $\lambda$ then effectively measures the collective direction of the QCD radiation,
while $x$ corresponds to its invariant mass.
Note that these new variables range in the interval $[0,1]$.
The potential leftover singularities in the $(\lambda,x)$ space are factorised 
and lie on the faces of the $[0,1]\times[0,1]$ square
without any extra divergences as the corners are approached.

With these definitions we find after some algebra (see for example~\cite{Anastasiou2013a})
the following parametrisation for the two particle massive phase space measure,
\begin{align}
\label{eq:HiggsInt}
d\Phi_{HX}=
\frac{s^{\frac{d}{2}-2}}{4 (2\pi)^{d-1}}
d\Omega_{d-2} d\lambda\,
\bar{z}^{d-3} (\lambda\tilde{\lambda})^{\frac{d}{2}-2} (1-x)^{\frac{d}{2}-2}
\theta(s) \theta(\bar{z}) \theta(\lambda) \theta(\bar\lambda)\theta(\bar{x}),
\end{align}
where the integral over the $(d-2)$-dimensional solid angle can be easily carried out using
\beq
\int d\Omega_d = \frac{2\pi^{\frac{d}{2}}}{\Gamma\left(\frac{d}{2}\right)}.
\eeq
The total integration measure of eq.~\eqref{eq:measure} then becomes
\begin{equation}
\label{eq:phiphi}
d\Phi_{H+m} = dx \frac{\bar z^2 \lambda\bar\lambda}{(1-\bar{z}\lambda x)^2}
d\Phi_{HX}(x,\lambda) d\Phi_m(x,\lambda).
\end{equation}
It should be noted here that our parametrisation of $\mu^2$ in terms of $x$ and $\lambda$
spoils the traditional factorisation of the phase space into a convolution over two independent phase spaces.
Instead, our phase space is now factorised into an iterative form.
First we perform the phase space integral over the QCD radiation phase space $d\Phi_m$,
obtaining a result as a function of $x$ and $\lambda$.
Afterwards we can perform the integral
over the Higgs phase space $d\Phi_{HX}$ in terms of $\lambda$
and finally over the additional parameter $x$.

The expression given in eq.~\eqref{eq:phiphi} is valid for 2 or more final state partons.
The cases of zero or one final state partons need to be addressed separately as they represent limiting cases of the general parametrisation above.
It is however straightforward to parametrise these cases explicitly.
For zero partons we find
\beq
d\Phi_{H+0} = \frac{2\pi}{s}\delta(\bar z).
\eeq
and for one final state parton we define
\beq
\label{eq:1pmeasure}
d \Phi_{H+1} = \frac{s^{\frac{d}{2}-2}}{4 (2\pi)^{d-2}} \bar z^{d-3} d\Omega_{d-2} d\lambda\,
(\lambda\bar\lambda)^{\frac{d}{2}-2}
\theta(s) \theta(\bar{z}) \theta(\lambda) \theta(\bar\lambda).
\eeq

Now that we have derived an explicit parametrisation of the phase space for the Higgs boson it is useful to make contact again between our integration variables 
and the actual properties of the Higgs boson.
Let $y$ be the rapidity of the Higgs boson in the partonic centre of mass frame,
which is related to $Y$ by
\begin{align}
Y = y_0 + y,
\qquad
y_0 = \frac{1}{2}\log\frac{x_1}{x_2}.
\end{align}
The partonic rapidity and transverse momentum of the Higgs boson are related to $\lambda$ and $x$ by
\begin{align}
y = \frac{1}{2}\log\frac{1-\bar{z}\tilde\lambda}{1-\bar{z}\lambda},
\qquad
p_T^2= \frac{m_h^2}{z}\bar{z}^2 (1-x)\lambda\tilde\lambda.
\end{align}
It is easy to see that the parametrisation obtained here corresponds to the choice of variables introduced in eq.~\eqref{eq:vardefpty}.
The differential partonic cross section required to compute the Higgs-differential hadronic cross section in eq.~\eqref{eq:xsdiffhad2}
is simply obtained by performing all integrations over the final state degrees of freedom,
except for the integrations over $x$ and $\lambda$.

We would like to remark that, although the phase space parametrisation in this section
was derived specifically for the case of Higgs boson production for definiteness,
it actually holds for any single particle colourless final state.
Moreover, extending the $(x, \lambda)$ parametrisation to the case of the production
of a colourless final state system composed of more particles,
the separation of the phase space and reverse unitarity
still work as discussed without the need of any further change.

\section{Computation of the Higgs-Differential Cross Section through NNLO}
\label{sec:computation}

In the previous section we established a framework for the computation of Higgs-differential cross sections.
In this section we explicitly compute the differential cross section for the production
of a Higgs boson via the gluon fusion mechanism through NNLO in the infinite top mass limit.

\subsection{Partonic Cross Sections for Gluon Fusion}
We compute the Higgs boson cross section in an approximation to the full Standard Model
where the top quark mass is considered to be infinite and internal top quark loops can be integrated out.
This leads to an effective field theory where the Higgs is directly coupled to gluons~\cite{Inami1983,Shifman1978,Wilczek1977,Spiridonov:1988md} through an effective operator
\beq
\mathcal{L} = \mathcal{L}_{QCD} - \frac{1}{4}C^0 G_{\mu\nu}G^{\mu\nu}h.
\eeq
Here, $\mathcal{L}_{QCD} $ is the QCD Lagrangian, $h$ is the Higgs boson field, $G_{\mu\nu}$ the gluon field strength
and $C^0$ the Wilson coefficient~\cite{Chetyrkin1997a,Chetyrkin:2005ia,Schroder:2005hy} that arises from matching the effective theory to the full Standard Model. 
The QCD Lagrangian contains $n_f$ massless quark fields with $n_c$ colours. 
In the following we will compute perturbative corrections in the strong coupling constant $\alpha_S$ through NNLO using this effective theory:
\beq
\label{eq:etadef}
\frac{1}{z}\frac{d^2 \hat{\sigma}_{ij}}{dx\,d\lambda}(z,x,\lambda,m_h^2)= (C^0)^2 \,\hat{\sigma}_0 \, \eta_{ij}(z,x,\lambda)
=(C^0)^2 \, \hat{\sigma}_0\sum_{k=0}^\infty\left(\frac{\alpha_S}{\pi}\right)^k \eta^{(k)}_{ij}(z,x,\lambda),
\eeq
where we normalised the coefficient functions $\eta_{ij}$ to the Born cross section
\beq
\hat{\sigma}_0=\frac{\pi}{8(n_c^2-1)}.
\eeq

Perturbative QCD calculations are plagued by the presence of ultraviolet and infrared divergencies
that appear during the computational steps and cancel for well defined observables. 
We regulate these divergencies by applying the framework of dimensional regularisation in the $\overline{\mathrm{MS}}$ scheme,
continuing the number of space time dimensions to $d=4-2\epsilon$. 
Ultraviolet finite observables are obtained by renormalising the parameters of the theory. 
The required redefinition of the strong coupling constant and of the Wilson coefficient are given by
\beq
\label{eq:renorm}
\alpha_S^0=\alpha_S(\mu^2) \left(\frac{\mu^2}{4\pi}\right)^{\epsilon}e^{\epsilon \gamma_E} Z_\alpha(\mu^2),
\qquad
C^0=C(\mu^2)Z_C(\mu^2),
\eeq 
where $\gamma_E$ is the Euler-Mascheroni constant.
For convenience we provide the well known factors $ Z_\alpha(\mu^2)$  and $Z_C(\mu^2)$ in appendix~\ref{sec:AppDef}.
The Wilson coefficient can be found in appendix~\ref{sec:APPWilson}.

At any order in perturbation theory we distinguish 6 different initial state configurations of partons:
\beq\bsp
\label{eq:vars}
g(p_1)+g(p_2) &\,\to  H(p_h)+X \\ 
q(p_1)+ g(p_2) &\,\to H(p_h) +X\\ 
g(p_1)+ q(p_2) &\,\to H(p_h) +X\\ 
q(p_1)+ \bar q (p_2) &\,\to  H(p_h) +X  \\
q(p_1)+ q (p_2) &\,\to  H(p_h) +X  \\
q(p_1)+ q^\prime (p_2) &\,\to  H(p_h) +X  
\esp\eeq
Here $g$, $q$, $\bar q$ and $q^\prime$ represent a gluon, a quark, an anti-quark and a quark with a different flavour respectively.
All other combinations of explicit (anti-)quark flavours and gluons can be obtained from the ones above.
$X$ represents a specific partonic final state.
The partonic coefficient function for the individual channels at order $n$ for $m$ final state partons is given by
\beq
\label{eq:partoneta}
\eta^{(n)}_{ij\rightarrow H+X}(z,x^\prime,\lambda^\prime)=\frac{N_{ij}}{2 m_h^2 \hat\sigma_0}\int d\Phi_{H+m}
\delta(x-x^\prime)\delta(\lambda - \lambda^\prime)\sum \mathcal{M}^{(n)}_{ij\rightarrow H+X},
\eeq
where $\sum \mathcal{M}^{(n)}_{ij\rightarrow H+X}$ is the coefficient of $\alpha_S^n$ in the coupling constant expansion
of the modulus squared of all amplitudes for partons $i$ and $j$ producing the final state $H+X$, summed over polarisations. 
The integration measure $d\Phi_{H+m}$ was defined in eq.~\eqref{eq:measure},
while the initial state dependent prefactors $N_{ij}$ are given by
\begin{align}
N_{gg}&=\frac{1}{4(1-\epsilon)^2(n_c^2-1)^2},\nonumber\\
N_{gq}&=N_{qg}=\frac{1}{4(1-\epsilon)(n_c^2-1)n_c},\\
N_{q\bar q}&=N_{qq}=N_{qq^\prime}=\frac{1}{4n_c^2}.\nonumber
\end{align}

For the differential cross section through NNLO we require amplitudes with up to two additional partons in the final state as well as contributions with up to two loops.
At any given order $\alpha_S^n$,  the sum of the number of loops and the number of final state partons is equal to $n$.
All purely virtual partonic cross sections are identical to those required for the computation of the inclusive Higgs boson cross section
that were obtain at two loops in refs.~\cite{Harlander2000,Gehrmann2005}.
At NLO we require tree level matrix elements with one additional parton in the final state (R).
At NNLO we require real-virtual (RV) matrix elements with one loop and one additional parton in the final state
and double-real (RR) matrix elements with two additional final state partons.

In order to compute the partonic coefficient functions we generate the necessary Feynman diagrams using QGRAF~\cite{Nogueira1993}. 
We then perform spinor, tensor and colour algebra in a private \texttt{C++} code based on \texttt{GiNaC}~\cite{Bauer2000},
a code based on \texttt{FORM}~\cite{Vermaseren2000} and a private \texttt{Mathematica} package.
The resulting  expressions represent the phase space and loop integrands for the partonic cross sections. 
Using the reverse unitarity framework discussed in section~\ref{sec:RU},
we treat loop and phase space integration on equal footing and use IBP~\cite{Tkachov1981,Chetyrkin1981} identities to express the partonic cross sections in terms of master integrals.
We then compute the master integrals explicitly, as discussed in section \ref{ssec:masters}.
Inserting them into the partonic matrix elements, we obtain the partonic cross sections as a Laurent expansion in the dimensional regulator.

Each of the contributions corresponding to different parton and loop multiplicities is separately infrared divergent as made manifest by explicit poles in the dimensional regulator. 
After summing up the contributions, some of the divergencies cancel by virtue of the KLN theorem. 
The remaining divergencies are absorbed by a suitable redefinition of the parton distribution functions,
\beq
f_i(x_i)=\left( \Gamma_{ij} \circ f_j^R \right)(x_i),
\eeq
where the convolution indicated with $\circ$ is defined by 
\beq
(f\circ g)(z) =\int_0^1 dx\,dy\,\delta(z-x y) f(x) g(y)=\int_z^1 \frac{dx}{x} f(x) g\left(\frac{z}{x}\right).
\eeq
The infrared counterterms $\Gamma_{ij}$ consist of convolutions of splitting functions $P_{ij}^{(n)}$ and can be derived from the DGLAP equation;
for reference we provide its explicit form in appendix~\ref{sec:AppDef}.
The remaining physical parton distribution functions $f^R_i$ are process independent and are extracted from measurements.

In inclusive calculations, it is often useful to employ the commutativity and associativity of the convolutions
to rewrite them such that the infrared counter term $\Gamma$ is convoluted with the partonic cross section before convoluting the result with the bare parton distributions.
The first convolution between the counter term and the partonic cross section can thus be performed analytically.
The required splitting functions were obtained in refs.~\cite{Vogt2004,Moch2004} and the convolutions were performed for example in refs.~\cite{Buehler2013,Hoschele2013}.
For the purposes of this work, it is impractical to perform these convolutions analytically, as they depend on the additional parameters $\lambda$ and $x$ due to momentum conservation.
We perform the convolution of the physical parton distribution functions and the infrared counter term therefore numerically. 
The infrared counter terms $\Gamma$ themselves contain poles in $\epsilon$.
In order to obtain the correct contributions to the finite part,
it is therefore necessary to compute the lower order partonic cross sections to higher than finite power in the dimensional regulator.

To obtain a finite fixed order differential cross section we expand the product of bare parton distribution functions $f_i$,
Wilson coefficient $C^0$ and partonic coefficient functions $\eta_{ij}$ and truncate the product at fixed order.
After the combination of all these contributions, we complete the computation of the Higgs-differential production cross section through NNLO eq.~\eqref{eq:xsdiffhad2}.
Extending the computation to one order higher in the strong coupling constant (N$^3$LO) will require as an input
the partonic cross sections computed to one order higher in the dimensional regulator. 
One of the main results of this work are the analytic expressions for all the necessary partonic cross sections through NNLO
up to and including order $\epsilon^1$, as required for an N$^3$LO computation. 
This result is available in electronic form in an ancillary file submitted together with this publication.

\subsection{Evaluation of Master Integrals}
\label{ssec:masters}

In this section we elaborate on the computation of the master integrals that serve as building blocks for our partonic cross section. 
To validate our results, we follow two different strategies.
The first is based on the method of differential equations~\cite{Kotikov1991,Gehrmann2000,Henn2013}
and is well established in the field of high order computations.
As both strategies lead to identical results we will not discuss this method in more detail.

The second approach is based on direct evaluation of phase space and loop integrals. 
Here, we discuss only master integrals involving at least one final state parton as purely virtual corrections are well known~\cite{Harlander2000,Gehrmann2005}.
At NLO the only master integral is simply given by the phase space measure eq.~\eqref{eq:1pmeasure}.

\subsubsection{Double-real master integrals}

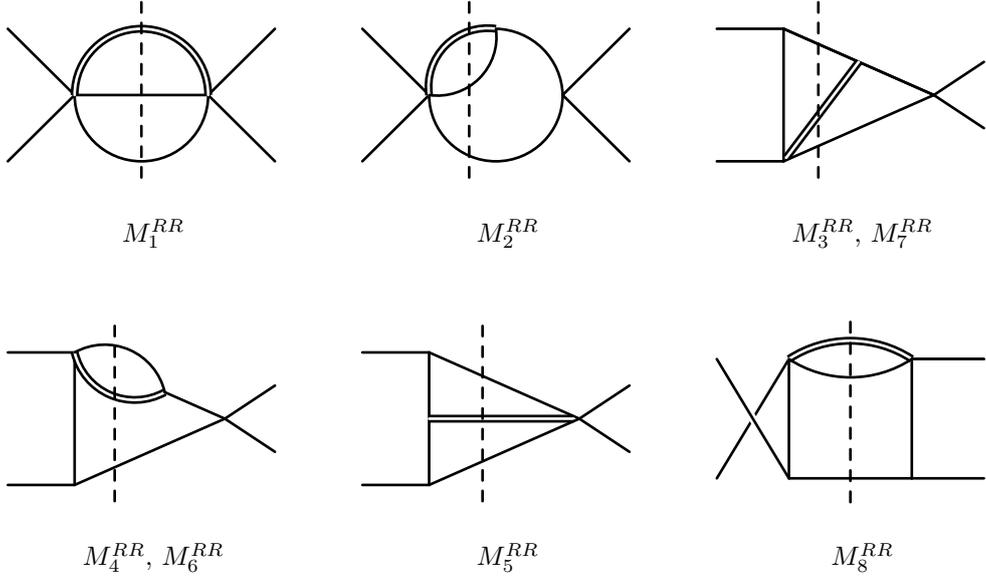
\begin{figure}[h!] 
\centering
\begin{subfigure}[c]{.30\linewidth}
\begin{minipage}[h][80pt][c]{120pt}
\centering
\begin{fmfgraph*}(100,50)
\fmfstraight
\fmfleft{i1,i2}\fmfright{o1,o2}
\fmf{plain,tension=2}{v2,o1}
\fmf{plain,tension=2}{v2,o2}
\fmf{plain,tension=2}{i1,v1}
\fmf{plain,tension=2}{i2,v1}
\fmf{plain,right}{v1,v2}
\fmf{plain,tension=0}{v1,v2}
\fmf{dbl_plain,left}{v1,v2}
\fmfforce{(.5w,1.2h)}{cut1}
\fmfforce{(.5w,-0.2h)}{cut2}
\fmf{dashes}{cut1,cut2}
\end{fmfgraph*}
\end{minipage}
\captionsetup{labelformat=empty}
\caption{$M_1^{RR}$}
\end{subfigure}
\begin{subfigure}[c]{.30\linewidth}
\begin{minipage}[h][80pt][c]{120pt}
\centering
\begin{fmfgraph*}(100,50)
\fmfstraight
\fmfleft{i1,i2}\fmfright{o1,o2}
\fmftop{t}
\fmf{plain,tension=2}{v2,o1}
\fmf{plain,tension=2}{v2,o2}
\fmf{plain,tension=2}{i1,v1}
\fmf{plain,tension=2}{i2,v1}
\fmf{plain,right}{v1,v2}
\fmf{phantom,left}{v1,v2}
\fmffreeze
\fmf{dbl_plain,left=.5,tension=0.5}{v1,t}
\fmf{plain,left=.4}{t,v2}
\fmf{plain,right=.5,tension=0.5}{v1,t}
\fmfforce{(.4w,1.2h)}{cut1}
\fmfforce{(.4w,-0.2h)}{cut2}
\fmf{dashes}{cut1,cut2}
\end{fmfgraph*}
\end{minipage}
\captionsetup{labelformat=empty}
\caption{$M_2^{RR}$}
\end{subfigure}
\begin{subfigure}[c]{.30\linewidth}
\begin{minipage}[h][80pt][c]{120pt}
\centering
\begin{fmfgraph*}(100,50)
\fmfstraight
\fmfleft{i1,i2}
\fmfforce{(1w,.25h)}{o1}
\fmfforce{(1w,.75h)}{o2}
\fmfforce{(.25w,1h)}{t}
\fmfforce{(.25w,0h)}{b}
\fmf{plain,tension=3}{v1,o1}
\fmf{plain,tension=3}{v1,o2}
\fmf{plain}{i1,b}
\fmf{plain}{i2,t}
\fmf{plain}{t,v1}
\fmf{plain}{v1,b}
\fmf{plain}{t,b}
\fmffreeze
\fmf{plain}{t,h}
\fmf{plain}{h,v1}
\fmffreeze
\fmf{dbl_plain}{m,h}
\fmf{phantom,tension=25}{m,b}
\fmfforce{(.38w,1.2h)}{c1}
\fmfforce{(.38w,-.2h)}{c2}
\fmf{dashes}{c1,c2}
\end{fmfgraph*}
\end{minipage}
\captionsetup{labelformat=empty}
\caption{$M_3^{RR}$, $M_7^{RR}$}
\end{subfigure}
\begin{subfigure}[c]{.30\linewidth}\vspace{2 em}
\begin{minipage}[h][80pt][c]{120pt}
\centering
\begin{fmfgraph*}(100,50)
\fmfstraight
\fmfleft{i1,i2}
\fmfforce{(1w,.25h)}{o1}
\fmfforce{(1w,.75h)}{o2}
\fmfforce{(.25w,1h)}{t}
\fmfforce{(.25w,0h)}{b}
\fmf{plain,tension=3}{v1,o1}
\fmf{plain,tension=3}{v1,o2}
\fmf{plain}{i1,b}
\fmf{plain}{i2,t}
\fmf{phantom}{t,v1}
\fmf{plain}{v1,b}
\fmf{plain}{t,b}
\fmffreeze
\fmf{plain,tension=1.5}{h,v1}
\fmf{phantom}{h,t}
\fmffreeze
\fmf{dbl_plain,right=.5}{t,h}
\fmf{plain,left=.5}{t,h}
\fmfforce{(.4w,1.2h)}{c1}
\fmfforce{(.4w,-.2h)}{c2}
\fmf{dashes}{c1,c2}
\end{fmfgraph*}
\end{minipage}
\captionsetup{labelformat=empty}
\caption{$M_4^{RR}$, $M_6^{RR}$}
\end{subfigure}
\begin{subfigure}[c]{.30\linewidth}\vspace{2 em}
\begin{minipage}[h][80pt][c]{120pt}
\centering
\begin{fmfgraph*}(100,50)
\fmfstraight
\fmfleft{i1,i2}
\fmfforce{(1w,.25h)}{o1}
\fmfforce{(1w,.75h)}{o2}
\fmfforce{(.25w,1h)}{b}
\fmfforce{(.25w,0h)}{t}
\fmfforce{(.25w,.5h)}{h}
\fmf{plain,tension=3}{v1,o1}
\fmf{plain,tension=3}{v1,o2}
\fmf{plain}{i1,t}
\fmf{plain}{i2,b}
\fmf{plain}{b,v1,t}
\fmf{plain}{t,b}
\fmffreeze
\fmf{phantom,tension=25}{m,v1}
\fmf{dbl_plain,tension=1}{h,m}
\fmfforce{(.45w,1.2h)}{c1}
\fmfforce{(.45w,-.2h)}{c2}
\fmf{dashes}{c1,c2}
\end{fmfgraph*}
\end{minipage}
\captionsetup{labelformat=empty}
\caption{$M_5^{RR}$}
\end{subfigure}
\begin{subfigure}[c]{.30\linewidth} \vspace{2 em}
\begin{minipage}[h][80pt][c]{120pt}
\centering
\begin{fmfgraph*}(100,45)
\fmfstraight
\fmfleft{i1,i2}\fmfright{o1,o2}
\fmfforce{(.27w,1h)}{t2}
\fmfforce{(.73w,1h)}{t3}
\fmfforce{(.27w,0h)}{b2}
\fmfforce{(.73w,0h)}{b3}
\fmf{plain,rubout=3pt}{i2,b2}
\fmf{plain}{i1,t2}
\fmf{plain}{t3,o2}
\fmf{plain}{b3,o1}
\fmf{plain}{t2,b2,b3,t3}
\fmf{plain}{o2,t3}
\fmffreeze
\fmf{dbl_plain,left=.3}{t2,t3}
\fmf{plain,left=-.3}{t2,t3}
\fmfforce{(.5w,-0.2h)}{cut1}
\fmfforce{(.5w,1.4h)}{cut2}
\fmf{dashes}{cut1,cut2}
\end{fmfgraph*}
\end{minipage}
\captionsetup{labelformat=empty}
\caption{$M_8^{RR}$}
\end{subfigure}
\caption{Double-real master integrals}
\label{figmaster}
\end{figure}

At NNLO there are contributions with two final state partons (double-real) as well as one-loop corrections to the emission of a single final state parton (real-virtual).
Let us consider the double-real radiation (RR) first.
We find eight master integrals $M_i^{RR}$, corresponding to the six diagrams in Figure~\ref{figmaster}.
Propagators crossing the dashed line in the diagrams represent cut propagators, while the double line marks the massive Higgs boson propagator.
The fact that we are interested in Higgs-differential master integrals means that the momentum of the Higgs boson is completely fixed,
so that no integration over the degrees of freedom of the Higgs boson occurs.
%
\begin{fmffile}{QWmasts}
\end{fmffile}
In accordance with the phase space definitions of eqs.~\eqref{eq:2mPS} and \eqref{eq:mPS},
all RR master integrals take then the form
\begin{equation}\label{2propint}
I(p,q;\alpha,\beta) \equiv \int \frac{d^dl}{(2\pi)^{d-2}} \frac{\delta(l^2)\delta((l-k)^2)}
{(l+p)^{2\alpha}(l+q)^{2\beta}}
\qquad
\begin{array}{l}
k^2\neq 0,\\p^2=0,\\ q^2\in\mathbb{R}
\end{array}
\end{equation}
where again we denoted with $k$ the momentum of the parton system. 
The generalised exponents $\alpha$ and $\beta$  and the momenta $q$ and $p$ specify the propagators that appear 
in the respective master integral. In particular, 
$p$ and $q$ are linear combinations of $p_1$, $p_2$ and $k$.
The integral over $l$ stands for the integral over one of the two final state parton momenta.
The result is then a function of all Lorentz invariants of the system.

The integral $I(p,q;\alpha,\beta)$ can be evaluated for generic values of the parameters
in terms of Appell's Hypergeometric function \cite{Somogyi:2011ir}:
\begin{multline}\label{angular integral}
I(p,q;\alpha,\beta)=(-1)^{\alpha+\beta}\frac{\pi ^{-1+\epsilon } 2^{-3-\beta +2 \epsilon } (k^2)^{-\epsilon }\Gamma (1-\beta -\epsilon)}
{\Gamma (2-\beta -2 \epsilon)}
\frac{(q^2-q\cdot k)^{\alpha-\beta}}{(k^2 (q\cdot p)-q^2 (k\cdot p))^{\alpha}}
\times\\\times
F_1\left(\alpha ;1-\beta-\epsilon,1-\beta -\epsilon;2-\beta -2 \epsilon
   ;1-\frac{1+\sqrt{1-4v_{11}}}{2 v_{12}},1-\frac{1-\sqrt{1-4v_{11}}}{2 v_{12}}\right)
\end{multline}
where we have defined the auxiliary quantities,
\begin{align*}
1-4v_{11}&
\equiv\frac{(q\cdot k)^2-k^2q^2}{(q^2-q\cdot k)^2}
\quad\xrightarrow{q^2=0} 1 ,
\\
2v_{12}&
\equiv\frac{q^2 (k\cdot p)-k^2 (q\cdot p)}{(p\cdot k)(q^2-q\cdot k)}
\quad\xrightarrow{q^2=0} 
\frac{k^2(q\cdot p)}{(p\cdot k)(q\cdot k)}.
\end{align*}
\\
The eight master integrals, shown in Figure~\ref{figmaster}, contributing to the RR partonic cross section, 
expressed in terms of the function $I(p,q;\alpha,\beta)$, are explicitly given by
\\
\begin{equation*} \label{mastersRR}
\begin{split}
M_1^{RR}&= C_{RR}(1-2 \epsilon) \, I(0,0;0,0), \\
M_2^{RR}&=-C_{RR}\frac{\zb \epsilon}{1-\lambda x \zb} \, I(0,p_1+p_2;0,1), \\
M_3^{RR}&= C_{RR}\frac{\lambda \zb \epsilon \left(\lambda ^2 x \zb^2-2 \lambda  x \zb+x \zb-\zb+1\right)} {1-\lambda x \zb} \, I(p_2,k-(p_1+p_2);1,1),\\
\end{split}\end{equation*}
\beq\begin{split}
M_4^{RR}&= C_{RR}(1-\lambda) \zb \epsilon \, I(p_1,p_1+p_2;1,1), \\
M_5^{RR}&=-C_{RR}\frac{(1-\lambda) \lambda  (1-x) \zb^2 \epsilon }{1-\lambda x \zb} \, I(p_2,k-p_1;1,1) ,\\
M_6^{RR}&= C_{RR}\frac{\lambda \zb \epsilon (1-x \zb)}{1-\lambda x \zb} \, I(p_2,p_1+p_2;1,1),\\
M_7^{RR}&= C_{RR}\frac{(1-\lambda) \zb \epsilon \left(\lambda ^2 \zb^2 x(1-x)-\zb+1\right)}{(1-\lambda x \zb)^2} \, I(p_1,k-(p_1+p_2);1,1),\\
M_8^{RR}&= C_{RR}\frac{(1-\lambda) \lambda x \zb^2 \epsilon }{1-\lambda x \zb} \, I(p_2,p_1;1,1),
\end{split}
\eeq
\\
where $M_6^{RR}$ and $M_7^{RR}$ are related respectively to $M_4^{RR}$ and $M_3^{RR}$ under the exchange of $p_1$ with $p_2$.
The rational prefactors and the $C_{RR}$ in the above definitions serve as a normalisation for the master integrals such that their expansion in the dimensional regulator
is given by a pure function of uniform transcendental weight.
This is true for all but $M_2^{RR}$ which still contains a root of a polynomial in our variables.
In particular, in the prefactor $C_{RR}$ we absorb additional factors that result from the phase space measure of the integration over the Higgs boson momentum, eq.~\eqref{eq:HiggsInt},
\beq
C_{RR}=\epsilon^3(\mu^2)^{2\epsilon} e^{2 \epsilon \gamma_E}\,
\frac{(4\pi)^{-2-\epsilon}\,(m_h^2)^{-\epsilon} z^\epsilon}{\Gamma(1-\epsilon)}
\zb ^{-2\epsilon}(\lambda \bar \lambda )^{-\epsilon}(1-x)^{-\epsilon}(1-\zb x \lambda)^{\epsilon}.
\eeq
We note that the number of differential master integrals is less than the number of inclusive master integrals found in the double real cross section in ref.~\cite{Anastasiou2002}.

\subsubsection{Real-virtual master integrals}

In the case of the RV master integrals, the integral over the final state parton phase space becomes trivial and all master integrals are simply given by one loop integrals. 
We define the two well known functions,
\begin{align}
\text{Bub}(p^2)&=\int \frac{d^dl}{i(2\pi)^d} \frac{1}{l^2(l+p)^2}\nonumber\\
&=\frac{(4 \pi )^{-2+\epsilon } (-p^2)^{-\epsilon } \Gamma (1-\epsilon )^2 \Gamma (\epsilon +1)}{\epsilon  (1-2 \epsilon ) \Gamma (1-2 \epsilon )},\nonumber\\
\text{Box}(q_1,q_2,q_3)&=\int \frac{d^dl}{i(2\pi)^d} \frac{1}{l^2(l+q_1)^2(l+q_1+q_2)^2(l+q_1+q_2+q_2)^2}.
\end{align}
The box integral was computed for example in ref.~\cite{Anastasiou2000a}.
With this we choose the following seven RV differential master integrals:
\begin{align}
\begin{split}
M_1^{RV}&=-\frac{2}{3}\epsilon (1-2\epsilon) \,C_{RV} \,\Re \left[\mathrm{Bub}\left((p_1+p_2)^2 \right)\right],\\
M_2^{RV}&=-\frac{2}{3}\epsilon (1-2\epsilon) \,C_{RV} \,\Re \left[\mathrm{Bub}\left(p_h^2\right)\right],\\
M_3^{RV}&=-\frac{2}{3}\epsilon (1-2\epsilon) \,C_{RV} \,\Re \left[\mathrm{Bub}\left((p_2-p_h)^2\right)\right],\\
M_4^{RV}&=-\epsilon^2s^2 \bar\lambda \zb \,C_{RV} \,\Re \left[\mathrm{Box}(p_2,p_1,-p_h) \right],\\
M_5^{RV}&=\epsilon^2 s^2 \lambda  \zb \,C_{RV} \,\left\{11 \zb \bar \lambda \Re \left[\mathrm{Box}(p_2,-p_h,p_1)\right]
- 23 \Re \left[\mathrm{Box}(p_1,p_2,-p_h)\right] \right\},\\
M_6^{RV}&=-\frac{2}{3}\epsilon (1-2\epsilon) \,C_{RV} \,\Re \left[ \mathrm{Bub}\left((p_1-p_h)^2\right)\right],\\
M_7^{RV}&=\epsilon^2 s^2 \lambda  \zb \,C_{RV} \,\left\{- \zb \bar \lambda \Re \left[\mathrm{Box}(p_2,-p_h,p_1)\right]
+ 2 \Re \left[\mathrm{Box}(p_1,p_2,-p_h)\right] \right \}.
\end{split}
\end{align}
The normalisation factor $C_{RV}$ again absorbs part of the integration measure of the Higgs boson momentum and is given by
\beq
C_{RV}=\epsilon^2 (\mu^2)^{2\epsilon}e^{2 \epsilon \gamma_E} \frac{(4 \pi)^{-1- \epsilon }\, (m_h^2)^{-\epsilon}\,z^\epsilon }{2\Gamma (1-\epsilon)}\,\zb^{-2 \epsilon} (\lambda \bar \lambda)^{-\epsilon}.
\eeq
Note that we include  the usual constants arising in the  $\overline{\mathrm{MS}}$ renormalisation procedure in the definition of $C_{RR}$ and $C_{RV}$.

\subsubsection{Expansion in the dimensional regulator}

For further numerical evaluation, we need to expand the master integrals as a Laurent series in the dimensional regulator $\epsilon$.
The Laurent expansions for the RV master integrals are well known, so we comment here only on the expansion of the RR integrals.
For masters $M_1$, $M_4$ and $M_5$, Appell's hypergeometric function can be reduced to Gauss' hypergeometric function ${}_2F_1$,
therefore one can obtain the $\epsilon$ expansion with well-known tools such as the \texttt{HypExp} package~\cite{Huber:2005yg,Huber:2007dx}.
For the remaining master integrals we can directly expand Appell's hypergeometric function through weight three, starting from the following integral representation:
\begin{align}\label{f1reps}
F_1(1;-\epsilon,-\epsilon;1-2\epsilon;x,y)&=
(-2\epsilon)\int_0^1 dt~ (1-t)^{-1-2\epsilon}(1-y t)^\epsilon(1-x t)^\epsilon
\nonumber\\
&=(-2\epsilon)\bar{x}^{\epsilon}\bar{y}^{\epsilon}\int_0^1dt~t^{-1-2\epsilon}\Big[\left(1-\left(1-tx'\right)^{\epsilon}\right)\left(1-\left(1-ty'\right)^{\epsilon}\right)
\nonumber\\
&\qquad-\left(1-\left(1-tx'\right)^{\epsilon}\right)-\left(1-\left(1-ty'\right)^{\epsilon}\right)\Big]
\nonumber\\
&+(-2\epsilon)\bar{x}^{\epsilon}\bar{y}^{\epsilon}\int_0^1dt \, t^{-1-2\epsilon}.
\end{align}
Here we have defined $x' = \frac{x}{x-1}$ and similarly for $y$. With this, the first integral in the second equality in eq.~\eqref{f1reps} is finite and can be expanded in $\epsilon$ before integrating. The divergence is captured by the second integral, which can be easily integrated for generic $\epsilon$. This yields the following expansion for the $F_1$:
\begin{eqnarray}
\label{eq:F1exp}
F_1(1;-\epsilon,-\epsilon;1-2\epsilon;x,y) &=& 1  + \epsilon \ln(\bar x \bar y)
-2 \epsilon^2 \left[
\Li_2(x) + \Li_2(y) + \frac{1}{4} \ln^2\left( \frac{\bar x}{ \bar y}\right)
\right] \nonumber \\ 
&& \hspace{-2.3cm}
-2 \epsilon^3 \left\{ 
2\,\Li_3(x) +2\, \Li_3(y)
-\Li_3\left( \frac{x}{y} \right)
+ \Li_3 \left( \frac{\bar x}{\bar y} \right)
+ \Li_3 \left( \frac {y \bar x}{x \bar y} \right) 
\right.\nonumber\\
&&-\ln  \left( \frac{\bar x}{\bar y} \right) \Li_2 \left(\frac {y \bar{x}}{x \bar{y}} \right) 
-\zeta_3 -\frac{1}{12} \ln^3 \left( \frac{\bar x}{\bar y} \right) 
\nonumber\\
&&\left.
-\frac{1}{6} \ln^3\left( -{\frac {x}{y}} \right) 
+\frac{1}{2} \ln(x) \ln^2 \left( \frac{\bar x}{\bar y} \right)
\right\}
+ {\cal O}( \epsilon^4).
\end{eqnarray}
The above expression is real-valued for $x \in [0,1], y<0$.

This allows us to express all coefficients in the Laurent series in terms of real valued classical polylogarithms, enabling a fast and stable numerical evaluation~\cite{Buehler2014}.


\subsection{Partonic Coefficient Functions}
\label{sec:partoniccoeffs}

In the previous section we obtained analytic results for all partonic coefficient functions required for the computation of the differential Higgs boson cross section through NNLO.
Moreover we computed the coefficient functions to sufficiently high order in the dimensional regulator such that the infrared subtraction term required for an N$^3$LO computation can be constructed.

It is important to note that our coefficient functions contain single poles in the variables $\zb$, $x$ and $\lambda$. 
These poles represent kinematic configurations where the Higgs boson degrees of freedom revert to lower multiplicity final state kinematics.
The prime example is a singularity in the matrix elements when the transverse momentum of the Higgs boson tends to zero, which is the value of the transverse momentum at leading order.
Specifically, these singularities are of the form
\beq
\left\{\zb^{-1+a_1 \epsilon},x^{-1+a_2 \epsilon},(1-x)^{-1+a_3 \epsilon},\lambda^{-1+a_4 \epsilon},(1-\lambda)^{-1+a_5 \epsilon}\right\}
\eeq
where the coefficients $a_i$ are small integer numbers. 

When integrating over the variables $\zb$, $\lambda$ and $x$ we encounter singularities
that lie at the boundaries of the integration range of our variables.
This is the case, for example, when we are computing observables like the rapidity distribution of the Higgs boson or the inclusive cross section.
In order to be able to compute such observables we will have to regulate the divergences, which we illustrate in the following example.

Consider a function $f(x) = x^{-1+a\epsilon}f_h(x)$, for some integer $a$ and with $f_h(x)$ holomorphic around $x=0$.
We are interested in integrating the function over a test function $\phi(x)$ on the range $[0,1]$.
In the case of our Higgs-differential cross section, the test function $\phi(x)$ corresponds to the product of the parton luminosity and the measurement function.
We can explicitly subtract the divergence at $x=0$ and integrate by parts to obtain
\bea
I&=&\int_0^1 dx f(x) \phi(x)= \int_0^1 x^{-1+a\epsilon} f_h(x) \phi(x)\nonumber\\
&=&\int_0^1 dx x^{-1+a\epsilon} \left[f_h(x)\phi(x)-f_h(0) \phi(0)\right] +\frac{1}{a\epsilon} f(0) \phi(0).
\eea
We now want to give an expression for the partonic cross section that is finite even if all inclusive integrations are performed.
To this end we define in a slight abuse of notation,
\beq
f_s(0)\equiv\delta(x)\left[x^{-1+a\epsilon}-\frac{1}{a \epsilon}\right]f_h(0).
\eeq
Here the $\delta$ distribution is to be understood as acting only on the test function and not on its coefficient in the square bracket.
It is easy to see that $f_s(0)$ integrates to zero.
We can therefore regulate the integrand $f(x)$ by subtracting $f_s(0)$,
\beq
I=\int_0^1 dx f(x) \phi(x)=\int_0^1 dx\,(f(x)- f_s(0))\,\phi(x),
\eeq
so that every term of its $\epsilon$ expansion can be integrated numerically.

In the case of our Higgs-differential cross sections, we need to regulate potential end-point divergences in the three remaining variables $\zb, x$ and $\lambda$,
c.f.~eq.~\eqref{eq:xsdiffhad2}.
We define the distributions $\sigma_s$ that subtract the limits of $\sigma(\zb, x, \lambda)$
and label them by the kinematic limit of the cross section that they reproduce.
For example $\sigma(\zb,0,\lambda)$ takes care of the limit of the cross section as $x$ goes to zero.
After partial fractioning to avoid simultaneous singularities on both endpoints of the integral, we obtain the following decomposition, 
\begin{align}
\label{eq:finitexs}
\sigma_f(\zb ,x,\lambda)\equiv & \,\sigma(\zb ,x,\lambda)-\sigma_s (\zb,x,1)-\sigma_s (\zb,x,0 )-\sigma_s (\zb,1,\lambda)-\sigma_s (\zb,0,\lambda)-\sigma_s (0,x,\lambda)\nonumber\\
+&\,\sigma_s (\zb,1,1)+\sigma_s (\zb,1,0)+\sigma_s (\zb,0,1 )+\sigma_s (\zb,0,0)+\sigma_s (0,x,1)+\sigma_s (0,x,0)\nonumber\\
+&\,\sigma_s (0,1,\lambda)+\sigma_s (0,0,\lambda )-\sigma_s (0,1,1)-\sigma_s (0,1,0)-\sigma_s (0,0,1)-\sigma_s (0,0,0).
\end{align}

One main result of this article is the analytic computation of the partonic coefficient functions $\eta^{(k)}_{ij}(z,x,\lambda)$ as defined in eq.~\eqref{eq:etadef}.
We created finite versions of this coefficient functions in the spirit discussed above
and provide them in \texttt{Mathematica} readable form in an ancillary file together with the arXiv submission of this article.
Specifically, we provide all 18 different kinematic configurations of these coefficient functions as in eq.~\eqref{eq:finitexs},
each as a sum of products of rational coefficients and master integrals.

\section{Numerical results for the Higgs diphoton signal}
\label{sec:numerics}
In this section, we carry out numerically the remaining integrations which are
necessary in order to obtain hadronic Higgs-differential cross sections.
We present results for the LHC at 13 TeV  in order to showcase
the types of observables which can be readily computed with our approach.

While our analytical results are original and extend the literature
of NNLO Higgs-differential cross sections at $\mathcal{O}(\epsilon)$,
our numerical predictions for the finite part can be compared to available computer codes.
We have validated our numerical implementation against the predictions
of HNNLO~\cite{Catani2007} and MCFM~\cite{Boughezal2016}
and we have found good agreement within Monte-Carlo uncertainties.

For our numerical studies we use NNLO MMHT parton
distribution functions~\cite{Harland-Lang2014} throughout,
as available from \texttt{LHAPDF}~\cite{Buckley:2014ana}.
Their default value for the strong coupling constant $\alpha_S(m_Z)=0.118$
and the corresponding three-loop running are also adopted.
We set the Higgs boson mass to $m_h=125$ GeV and neglect its width.
We equate the renormalisation and factorisation scales for simplicity
and we choose $\mu=m_h/2$ as a central value.
As it is common practice, we estimate the effect of missing higher order
corrections by varying $\mu$ by a factor of two around its central value.

\begin{figure*}[t]
\centering
\begin{subfigure}[b]{0.45\textwidth}
\includegraphics[width=0.95\textwidth]{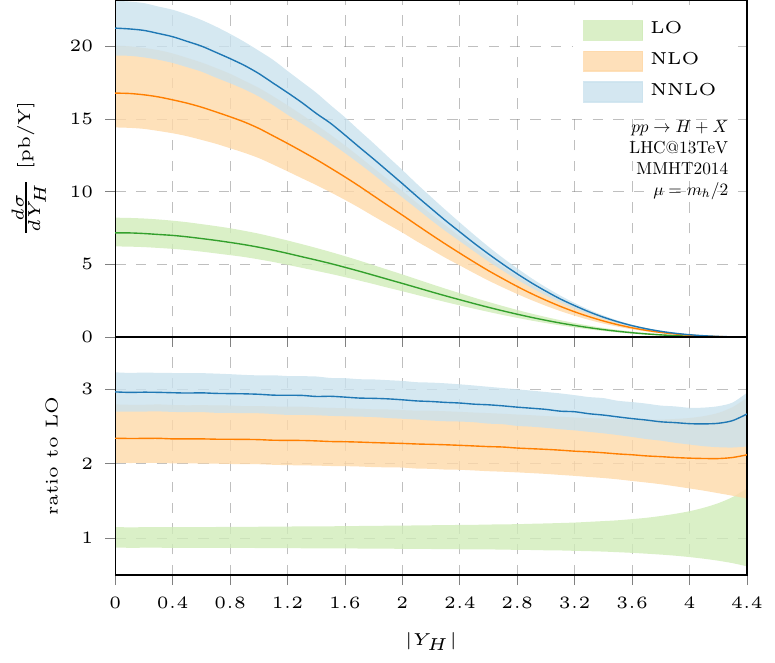}
\caption{
}
\label{fig:rapPhotInc}
\end{subfigure}
\begin{subfigure}[b]{0.45\textwidth}
\includegraphics[width=0.95\textwidth]{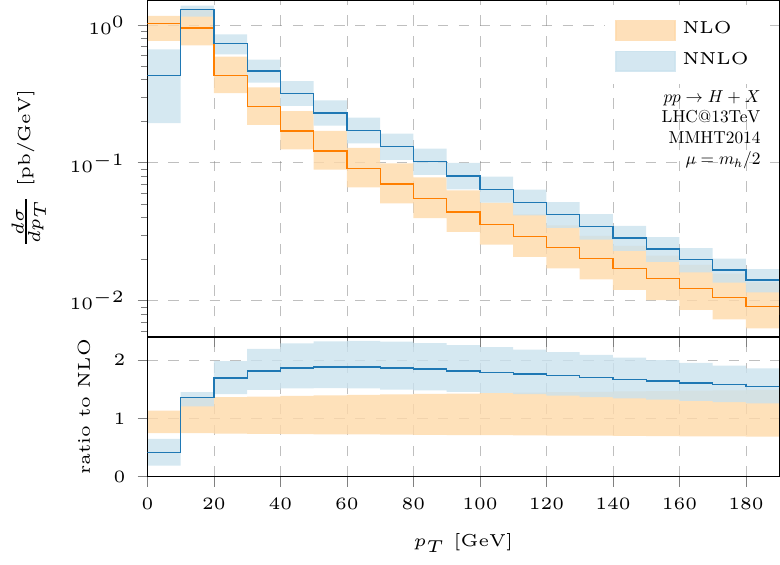}
\caption{
}
\label{fig:pt_dist_inc}
\end{subfigure}
\caption{
  Rapidity and $p_T$ distributions for the inclusive production of a Higgs boson via gluon fusion for different orders in perturbative QCD through NNLO.
  The bands represent the variation of the cross section under changes of the unphysical scale $\mu$ w.r.t. the central value (thick lines) by a factor of two.
}
\end{figure*}

We start by showing distributions for the inclusive production of a Higgs boson via gluon fusion.
In Figure~\ref{fig:rapPhotInc} we present the unbinned rapidity distribution of the Higgs boson.
The bands correspond to the variation of the cross section at LO, NLO and NNLO in our default $\mu$ scale range.
In Figure~\ref{fig:pt_dist_inc} we show the $p_T$ distribution of the Higgs boson.
In order to have a non-vanishing transverse momentum, 
the Higgs boson needs to recoil against additional radiation
and therefore its $p_T$ distribution is trivial at LO.

In addition to simple inclusive distributions for a stable Higgs boson, 
we can also investigate its decays.
Specifically, we present differential distributions for the Higgs
diphoton signal after the application of typical selection cuts for
the photons.  The decay of the Higgs boson to two photons allows for
precise measurements  of numerous properties (see
e.g.~\cite{Tavolaro,ATLASCollaboration2015,ATLASCollaboration2014,ATLASCollaboration2014a,CMSCollaboration2014,ATLASCollaboration2013,TheAtlasCollaboration2016})
due to its exceptionally clean experimental signature and it has played
a crucial role in the discovery of the Higgs boson  itself~\cite{Aad2012,Chatrchyan2012}.

We impose photon selection cuts which follow as closely as possible
the diphoton analysis of ATLAS~\cite{TheAtlasCollaboration2016}.
We require that the pseudorapidities of both photons satisfy $\eta_{\gamma} < 2.37$,
together with $\eta_\gamma \not\in [1.37,1.52]$, 
which implies that there are no photons in these regions. 
Furthermore, we require the photon with
larger transverse momentum to satisfy $p_{T,\,\gamma_1}>0.35 \,m_h$
and the one with smaller transverse momentum to have
$p_{T,\,\gamma_2}>0.25 \, m_h$. Since we have integrated out the
associated radiation in the production of the Higgs boson, our
analysis ignores photon isolation\footnote{This is an important cut for taming
the diphoton background but it is a minor one for the Higgs signal process that
we study here.}.\skip\footins=\bigskipamount

\begin{figure}[!ht]
\centering
\includegraphics[width=0.5\textwidth]{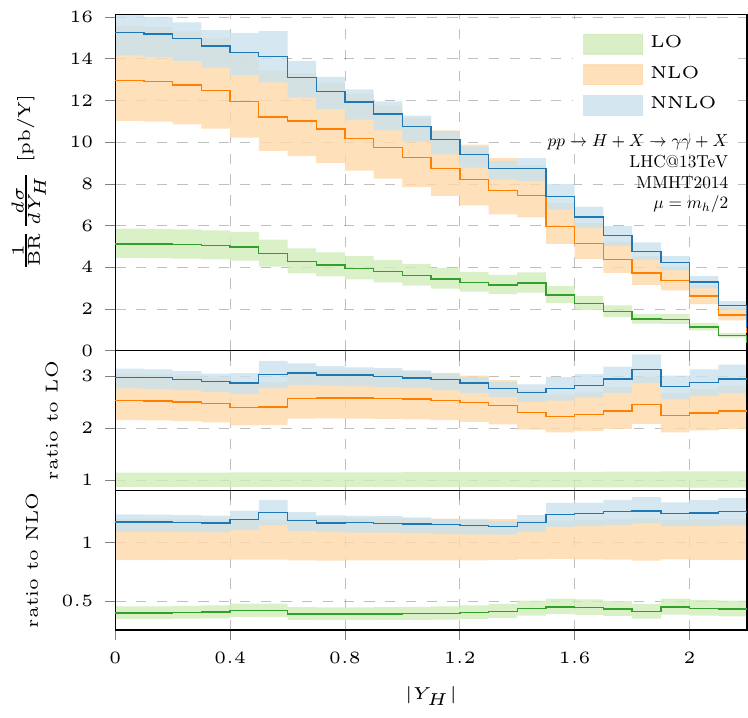}
\caption{
The Higgs boson rapidity for different orders in perturbative QCD through NNLO in the fiducial volume for the diphoton decay.
The bands represent the variation of the cross section under changes of the unphysical scale $\mu$ w.r.t. the central value (thick lines) by a factor of two.
}
\label{fig:rapPhot}
\end{figure}

In Figure~\ref{fig:rapPhot},  we present the rapidity distribution of
the Higgs boson after the application of the photon selection cuts described above.
The bands correspond to the values of the cross section at LO,
NLO and NNLO in our default $\mu$ scale range.
Due to the restrictions in the coverage of the pseudorapidity of the photons,
the Higgs rapidity distribution manifests some non-smooth changes.
In the middle panel we show the conventional $K$-factors,
i.e.\ we normalise the rapidity distribution at LO, NLO and NNLO for a general scale $\mu$
to the LO prediction evaluated at a fixed scale $\mu = m_h/2$.
We observe that the relative size of QCD corrections at NLO and NNLO
has a pattern which is similar to the one seen for the inclusive cross section,
although the $K$-factors are not entirely uniform across bins
due to the effect of the photon cuts.
In the lower panel, we normalise the rapidity distribution at LO, NLO and NNLO for a general scale $\mu$
to the NLO prediction at NLO at a fixed scaled $\mu = m_h/2$.
This shows that the relative $K$-factor from NLO to NNLO is more uniform in rapidity.
\begin{figure*}[!ht]
\centering
\begin{subfigure}[t]{0.45\textwidth}
\includegraphics[width=0.95\textwidth]{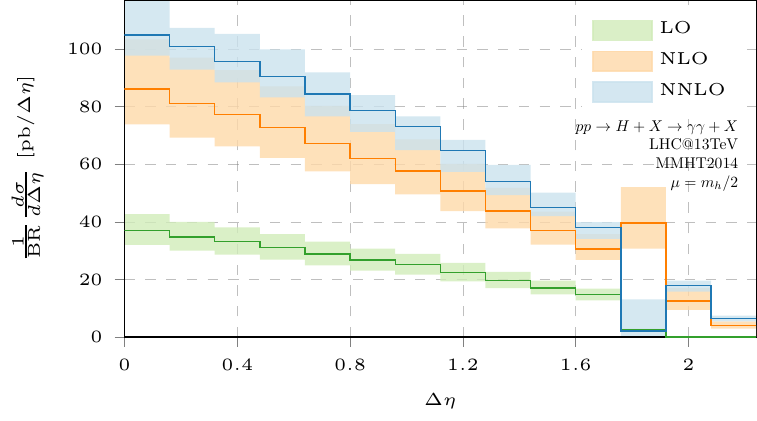}
\caption{Pseudorapidity difference of the two final state photons, $\Delta \eta=|\eta_{\gamma_1}-\eta_{\gamma_2}|$.}
\label{fig:PhotonRap}
\end{subfigure}
\quad
\begin{subfigure}[t]{0.45\textwidth}
\includegraphics[width=0.95\textwidth]{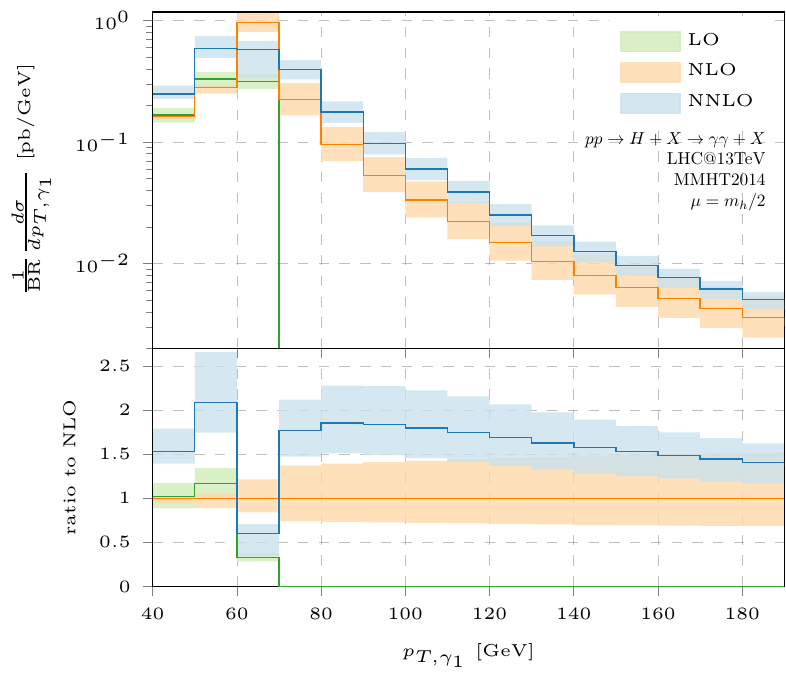}
\caption{Distribution of the largest photon transverse momentum $p_{T,\,\gamma_1}$.}
\label{fig:PhotonPT}
\end{subfigure}
\caption{
Differential distributions for the decay products of the Higgs boson for different orders in perturbative QCD through NNLO in the fiducial volume for the diphoton decay.
The bands represent the variation of the cross section under changes of the unphysical scale $\mu$ w.r.t. the central value (thick lines) by a factor of two.
}
\end{figure*}
In Figure~\ref{fig:PhotonRap}, we present the distribution of the pseudorapidity difference of the two photons.
The distribution has a kinematic edge at leading order at $\Delta \eta \simeq 1.79$. 
Above this point it features a Sudakov shoulder:
fixed order perturbative corrections are not trustworthy and resummation is required.
However, the bulk of the distribution can be calculated in fixed order perturbative QCD. 

In Figure~\ref{fig:PhotonPT}, we present the $p_T$ distribution of the leading photon.
At LO, the photon $p_T$ cannot exceed the value of $m_h/2$.
In addition, the experimental selection imposes a lower $p_T$ value at $0.35 \,m_h$.
This severe restriction of the phase space 
leads to large corrections to the available perturbative results.
As such, resummation would be required to obtain stable predictions in this kinematical regime.
Beyond LO the phase space for larger values of $p_T$ opens up and the distribution becomes more well-behaved beyond 100 GeV.

These distributions are just a small selection of possible observables that can be computed in our framework.
Combining the production with other decay modes is straightforward and can be used to study a number of phenomenologically relevant observables.

\section{Conclusion}
\label{sec:conclusions}
In this article, we presented differential distributions for Higgs boson observables at NNLO in perturbative QCD,
obtained within our ``Higgs-differential'' framework.
Our computational method is a departure from common frameworks for differential calculations
in that it avoids explicit subtraction of infrared divergences at the cost of being inclusive in jet observables.
This is achieved by separating the QCD radiation phase space from the phase space of the produced colour-neutral final state.
By integrating the QCD radiation phase space inclusively in dimensional regularisation, soft and collinear divergences are made explicit as poles in the regulator.
Furthermore, this separation enables us to employ reverse unitarity and related techniques that have been developed in inclusive calculations,
in order to simplify the phase-space integrations over final-state partons which are produced in association with the Higgs boson.

The NNLO cross section has been cast in terms of few master integrals, which we compute in an arbitrary number of
dimensions in terms of standard hypergeometric functions that admit expansions to arbitrarily high order in the dimensional regulator.
We have presented results that go beyond the finite term in the $\epsilon$ expansion, which were unknown in the literature.
These new results are essential ingredients for the construction of collinear and UV counter terms for Higgs-differential cross sections at N$^3$LO.
The master integrals encountered here do not depend on the exact nature of the colourless final state produced
and could be directly re-used in a similar differential Drell-Yan calculation.

Finally, we implemented the Higgs boson cross section through NNLO in a numeric code
and tested its efficiency in kinematic distributions for the Higgs boson and its decay products in the diphoton signal.
The predictions for these distributions were compared against results obtained from existing Monte-Carlo generators, thus validating our approach.

The main motivation for the approach presented in this article is its extensibility to even higher orders in QCD perturbation theory.
The general framework of separating the QCD radiation phase space from the Higgs phase space persists at higher orders.
This completely isolates the numerical calculation of distributions from the complications of higher order QCD computations.
These should only be reflected in the master integrals that appear at higher orders.
In this respect, it has been encouraging that the NNLO master integrals for Higgs-differential cross sections are relatively simple to compute.
Using techniques that have been honed in inclusive calculations at N$^3$LO, it should therefore be possible to compute the required master integrals,
leading to differential predictions for a hadron collider at N$^3$LO in QCD perturbation theory.

\section{Acknowledgement}
We thank Babis Anastasiou for inspiring discussions and useful comments on the manuscript.
We are also grateful to Simone Alioli, Claude Duhr and Achilleas Lazopoulos for fruitful discussions
and to Alexander Huss for numerical comparisons and valuable exchange of views.
SL, AP and CS are supported by the ETH Grant ETH-21 14-1 and the Swiss National Science
Foundation (SNSF) under contracts 165772 and 160814.
The work of FD is supported by the U.S. Department of Energy (DOE) under contract DE-AC02-76SF00515.
BM is supported by the European Commission through the ERC grants ``pertQCD'' and ``HICCUP''.

\appendix
\section{Renormalization Factors}
\label{sec:AppDef}
The renormalisation factors for the strong coupling constant and Wilson coefficient~\eqref{eq:renorm} required for a computation through N$^3$LO~\cite{Gehrmann2010} are given by
\bea
Z_\alpha&=&1+\frac{\alpha_S}{\pi}\left(-\frac{\beta_0}{\epsilon }\right)+\left(\frac{\alpha_S}{\pi}\right)^2\left(\frac{\beta_0^2}{\epsilon ^2}-\frac{\beta_1}{2 \epsilon }\right)+\left(\frac{\alpha_S}{\pi}\right)^3\left(-\frac{\beta_0^3}{\epsilon ^3}+\frac{7 \beta_1 \beta_0}{6 \epsilon ^2}-\frac{\beta_2}{3 \epsilon }\right)+\mathcal{O}(\alpha_S^4).\nonumber\\
Z_C&=&1-\frac{\alpha_S}{\pi}\left(\frac{\beta_0}{\epsilon}\right) +\left(\frac{\alpha_S}{\pi}\right)^2\left(\frac{\beta_0^2}{\epsilon^2}-\frac{\beta_1}{\epsilon}\right) 
-\left(\frac{\alpha_S}{\pi}\right)^3\left(\frac{\beta_0^3}{\epsilon^3}-\frac{2\beta_0 \beta_1}{\epsilon^2}+\frac{\beta_2}{\epsilon}\right) +\mathcal{O}(\alpha_S^4).\nonumber\\
\eea 
The coefficients at the various orders in the coupling constant $\beta_i$ are given by the QCD beta function~\cite{VanRitbergen1997,Czakon2005,Baikov2016,Herzog2017}.

The infrared counter term $\Gamma$ consists of convolutions~\cite{H??schele2013}  of splitting functions $P_{ij}^{(n)}$ and can be derived from the DGLAP equation. 
Its perturbative expansion required for an N$^3$LO accurate calculation of the differential Higgs boson production cross section is given by
\bea
\Gamma_{ij}&=&\delta_{ij}\delta(1-x)\nonumber\\
&+&\left(\frac{\alpha_S}{\pi}\right)\frac{P^{(0)}_{ij}}{\epsilon}\nonumber\\
&+&\left(\frac{\alpha_S}{\pi}\right)^2\left[\frac{1}{2\epsilon^2}\left(P^{(0)}_{ik}\circ P^{(0)}_{kj}-\beta_0  P^{(0)}_{ij}\right)+\frac{1}{2\epsilon}P^{(1)}_{kj}\right]\\
&+&\left(\frac{\alpha_S}{\pi}\right)^3\left[\frac{1}{6\epsilon^3}\left(P^{(0)}_{ik}\circ P^{(0)}_{kl}\circ P^{(0)}_{lj}-3\beta_0P^{(0)}_{ik}\circ P^{(0)}_{kj}+2\beta_0^2P^{(0)}_{ij}\right)\right.\nonumber\\
&&\left.+\frac{1}{6\epsilon^2}\left(P^{(1)}_{ik}\circ P^{(0)}_{kj}+2P^{(0)}_{ik}\circ P^{(1)}_{kj}-2\beta_0 P^{(1)}_{ij}-2 \beta_1 P^{(0)}_{ij}\right)+\frac{1}{3\epsilon}P^{(2)}_{ij}\right].\nonumber
\eea

\section{Wilson Coefficient}
\label{sec:APPWilson}
In the effective theory with $n_f$ light flavours and the top quark decoupled from 
the running of the strong coupling constant,
the $\overline{\textrm{MS}}$-scheme Wilson coefficient 
 reads~\cite{Chetyrkin:1997un,Schroder:2005hy}
\begin{eqnarray}
C(\mu^2) & = & - \frac{\alpha_S}{3 \pi v} \Bigg\{ 1 +\left(\frac{\alpha_S}{\pi} \right)\,\frac{11}{4} 
+ \left(\frac{\alpha_S}{\pi} \right)^2 \left[\frac{2777}{288} - \frac{19}{16} \log\left(\frac{m_t^2}{\mu^2}\right) -
 n_f\left(\frac{67}{96}+\frac{1}{3}\log\left(\frac{m_t^2}{\mu^2}\right)\right)\right] 
 \nonumber \\
  &&+ \left(\frac{\alpha_S}{\pi} \right)^3\Bigg[-\left(\frac{6865}{31104} 
  + \frac{77}{1728} \log\left(\frac{m_t^2}{\mu^2}\right) 
  + \frac{1}{18}\log^2\left(\frac{m_t^2}{\mu^2}\right)\right) n_f^2 \\
  && + \left(\frac{23}{32} \log^2\left(\frac{m_t^2}{\mu^2}\right) - 
  \frac{55}{54} \log\left(\frac{m_t^2}{\mu^2}\right)+\frac{40291}{20736} 
 - \frac{110779}{13824} \zeta_3 \right) n_f \nonumber\\
  &&  -\frac{2892659}{41472}+\frac{897943}{9216}\zeta_3 
  + \frac{209}{64} \log^2\left(\frac{m_t^2}{\mu^2}\right)
  - \frac{1733}{288}\log\left(\frac{m_t^2}{\mu^2}\right)\Bigg] +\mathcal{O}(\alpha_S^4) \Bigg\} \, .\nonumber
\end{eqnarray}
\bibliography{biblio}
\bibliographystyle{JHEP}

\end{document}